\documentclass[aps, prd, twocolumn, nofootinbib, superscriptaddress, floatfix, preprintnumbers]{revtex4}
\usepackage{amsmath, amssymb, amsfonts, mathtools}
\usepackage{bm, slashed, physics, mathrsfs, cases}
\usepackage{epsfig, graphicx}
\usepackage{tabularx, multirow, hhline, float}
\usepackage{color}
\usepackage[dvipsnames]{xcolor}
\usepackage[colorlinks=true, citecolor=blue, linkcolor=blue, urlcolor=blue, anchorcolor=blue, filecolor=blue, menucolor=blue, runcolor=blue]{hyperref}
\usepackage[normalem]{ulem}
\usepackage{orcidlink}
\usepackage{appendix}

\newcommand{\non}{\nonumber}

\begin{document}

\title{Towards Systematic Control of the Nucleon Gluon PDF with the Pseudo-PDF Method}

\author{Alex NieMiera \orcidlink{0009-0007-0553-2590}}
\email{niemiera@msu.edu}
\affiliation{Department of Physics and Astronomy, Michigan State University, East Lansing, Michigan 48824, USA}
\affiliation{Department of Computational Mathematics,
  Science and Engineering, Michigan State University, East Lansing, Michigan 48824, USA}

\author{Kai Ebira \orcidlink{0009-0007-1042-5271}}
\email{ebirakai@msu.edu}
\affiliation{Department of Physics and Astronomy, Michigan State University, East Lansing, Michigan 48824, USA}
\affiliation{Department of Computational Mathematics,
  Science and Engineering, Michigan State University, East Lansing, Michigan 48824, USA}
  
\author{William Good \orcidlink{0000-0001-8428-1003}}
\email{goodwil9@msu.edu}
\affiliation{Department of Physics and Astronomy, Michigan State University, East Lansing, Michigan 48824, USA}
\affiliation{Department of Computational Mathematics,
  Science and Engineering, Michigan State University, East Lansing, Michigan 48824, USA}

\author{Huey-Wen Lin \orcidlink{0000-0001-6281-944X}}
\affiliation{Department of Physics and Astronomy, Michigan State University, East Lansing, Michigan 48824, USA}

\begin{abstract}
    We present updated lattice quantum chromodynamics (QCD) results for the unpolarized nucleon gluon parton distribution function (PDF) using the pseudo-PDF methodology. 
    The calculation is performed at valence pion masses $M_\pi \approx 310$ and $690$ MeV on an $a \approx 0.06$ fm ensemble with $2+1+1$ flavors of highly improved staggered quarks (HISQ) generated by the MILC Collaboration. This lattice spacing is finer than those used in previous gluon pseudo-PDF studies, providing access to multiple short-distance separations and nucleon boost momenta as high as $4.4$ GeV. 
    To control excited-state contamination and reduce model dependence, we employ the recently developed Lanczos method, finding good agreement with our previous two-state fit approach. Higher-twist contributions and discretization effects are accounted for through explicit parameterizations in fits to the renormalized short-distance matrix elements. 
    Together, these advances represent important steps toward systematically controlled, high-precision extractions of the gluon PDF. 
    The resulting PDFs exhibit no statistically significant dependence on the valence pion mass and indicate a strongly suppressed gluon distribution at large $x$, consistent with a subset of phenomenological PDF extractions. 
    These results demonstrate progress toward providing systematically controlled lattice-QCD constraints for future global PDF analyses.
    
\end{abstract}

\preprint{MSUHEP-26-018}
\date{\today}
\maketitle

\section{Introduction}\label{sec:intro}

Quantum chromodynamics (QCD), the non-Abelian gauge theory of quarks and gluons, derives much of its richness and complexity from the gluon sector, whose self-interactions distinguish it from the relative simplicity of electrodynamics. 
Although absent from the simplest valence-quark picture of the nucleon, gluons play a central role in determining its internal structure. 
They carry approximately half of the nucleon's longitudinal momentum at typical high-energy reference scales and are intimately connected to confinement and the origin of visible mass in the universe~\cite{Accardi:2012qut,Gross:2022hyw}. 
The leading-twist, unpolarized gluon parton distribution function (PDF), $g(x,Q^2)$, characterizes the distribution of gluon longitudinal momentum fractions $x$ within the nucleon at a resolution scale $Q^2$. 
Understanding this distribution is therefore essential to explicating the role of gluons in nucleon structure, including their contributions to its mass and spin as described by matrix elements of the QCD energy-momentum tensor~\cite{Ji:1994av,Jaffe:1989jz}.

Despite its importance, the gluon PDF remains a challenging quantity determine precisely. 
Since gluons do not couple directly to the virtual photon in electromagnetic deep-inelastic scattering (DIS), their distribution is constrained indirectly through the $Q^2$ evolution of quark structure functions and through complementary measurements of gluon-sensitive processes, including inclusive jet, heavy-quark, and prompt-photon production~\cite{Accardi:2016ndt,Rojo:2015acz}.
Furthermore, accurate knowledge of the gluon PDF is particularly important for hadron-collider phenomenology, where it governs predictions for processes such as Higgs boson production through gluon fusion, top-antitop pair production, and high-energy dijet production. 
Uncertainties in the gluon distribution consequently propagate into precision tests of the Standard Model and searches for new physics~\cite{Czakon:2013tha,AbdulKhalek:2018rok}.
These uncertainties are especially pronounced at small $x$, where the gluon distribution rises rapidly and nonlinear phenomena such as gluon saturation may become relevant, and at large $x$, where experimental constraints remain limited. 
Improving the determination of the gluon PDF is thus a continuing priority for both experimental and theoretical studies~\cite{Hou:2019efy,Ball:2017nwa,AbdulKhalek:2021gbh,Achenbach:2023pba,Amoroso:2022eow,Ethier:2020way,Burkert:2022hjz,Proceedings:2026xrb}.

Within the last decade, lattice QCD has provided a complementary, first-principles approach to determining parton distributions. 
In particular, large-momentum effective theory (LaMET)~\cite{Ji:2013dva} and the pseudo-PDF method~\cite{Balitsky:2019krf} have enabled access to the longitudinal momentum dependence of PDFs through spatially separated operators evaluated in Euclidean space-time.
While these approaches have yielded substantial progress in the determination of quark distributions, gluon PDFs remain comparatively difficult to extract due to the poor signal-to-noise ratio of gluonic matrix elements.
Nevertheless, pseudo-PDF calculations have established a growing body of results for the gluon distributions of the pion~\cite{Fan:2021bcr}, kaon~\cite{Salas-Chavira:2021wui,NieMiera:2025inn}, and nucleon~\cite{Fan:2022kcb,Good:2023gai}.
More recently, advances in lattice calculations have also enabled the first extractions of gluon PDFs using LaMET~\cite{Good:2024iur, Good:2025daz,NieMiera:2025mwj, NieMiera:2025vcx}. 
These developments demonstrate the growing potential of lattice QCD to constrain gluon structure, while highlighting the need for improved systematic control at various stages of the analyses.

In particular, the extraction of nucleon gluon PDFs requires careful control of discretization effects, excited-state contamination, and systematic uncertainties in the PDF reconstruction. 
These challenges are particularly pronounced for gluonic matrix elements, which are disconnected from the valence--quark lines of the nucleon interpolating fields and are extracted from correlations between the nucleon two-point function and a gluonic operator containing a spatial Wilson line.
The resulting signal-to-noise ratio deteriorates with increasing source-sink separation and nucleon boost momentum, limiting the Euclidean time range over which statistically reliable data can be obtained. 
Excited-state contamination can therefore remain significant, complicating the isolation of the ground-state matrix element.
Moreover, conventional multi-state fits can exhibit sensitivity to the chosen fit window and number of states, making robust control of these effects essential for reliable PDF extraction. 
In this work, we address these challenges through a combination of fine lattice spacing, explicit control of excited-state contamination, and systematic treatment of the PDF reconstruction.
To constrain excited-state effects, we employ the Lanczos method~\cite{Lanczos:1950zz}, recently applied to a variety of lattice QCD spectroscopy and matrix elements extractions~\cite{Wagman:2024rid,Hackett:2024xnx,Ostmeyer:2024qgu,Abbott:2025yhm,Detmold:2026vjx,Zhang:2025hyo,Grebe:2026qmt}. 
Building on these developments, we combine this excited-state control of fine-lattice-spacing matrix elements and a systematic treatment of PDF extraction using parameterizations previously used in quark pseduo-PDF extractions~\cite{Karpie:2021pap, JeffersonLabAngularMomentumJAM:2022aix} to assess the reliability of the resulting nucleon gluon PDFs.
The remainder of this paper is organized as follows. 
In Sec.~\ref{sec:details}, we describe the lattice setup and matrix-element analysis, including our implementation of the Lanczos method. 
In Sec.~\ref{sec:results}, we present the gluon PDF extraction and discuss our results. 
Finally, in Sec.~\ref{sec:conclusion}, we summarize our findings and outline directions for future improvements.

\section{Lattice Details and Analysis Method}
\label{sec:details}
We carry out a lattice QCD calculation of the unpolarized gluon PDF of the nucleon at a physical pion mass on an $N_f = 2+1+1$ ensemble of highly improved staggered quarks (HISQ)~\cite{Follana:2006rc} generated by the MILC Collaboration~\cite{MILC:2010pul,MILC:2012znn,MILC:2015tqx}.
The ensemble has a lattice spacing of $a \approx 0.0582(4)$~fm and a volume of $L^3 \times T = 48 \times 144$.
The results presented in this work are obtained using $676,512$ nucleon two-point correlator measurements.
We use Wilson--clover fermions in the valence sector, with the valence-quark masses tuned to reproduce the masses of the light and strange pseudoscalar mesons in the sea sector, corresponding to pseudoscalar masses of approximately $~310$ and $~690$ MeV, respectively. 
Such a mixed-action setup has been extensively studied in previous works~\cite{Mondal:2020cmt,Park:2020axe,Jang:2019jkn,Jang:2019vkm,Gupta:2018lvp,Lin:2018obj,Gupta:2018qil,Gupta:2017dwj,Bhattacharya:2015wna,Bhattacharya:2015esa,Bhattacharya:2013ehc,Bhattacharya:2011qm,Briceno:2012wt,Yoon:2016jzj,Bhattacharya:2016zcn} and found to be free of exceptional configurations, which could otherwise render the Dirac operator ill-conditioned or lead to anomalously large correlation functions.

The nucleon two-point (2pt) correlation functions are computed on each gauge configuration using multiple source locations and are defined as
\begin{equation} \label{eq:2pt-correlator}
    C_{\rm N}^{\rm 2pt} (P_z;t)
    =
    \int d^3 y \; e^{-i \mathbf{y}\cdot \mathbf{\hat{z}}P_z} \langle 0 | \chi(\vec{y}; t) | \bar{\chi}(\vec{0}; 0) | 0 \rangle,
\end{equation}
where $P_z$ denotes the nucleon boost momentum along the $\hat{z}$-direction and $t$ is the Euclidean time separation between the source and sink. 
The nucleon interpolating operator is given by $\chi = \epsilon_{\rm abc}(d_{\rm a}^T C \gamma_5 \Gamma_\text{NR} u_{\rm b}) \mathcal{P}_+ u_{\rm c}$, where $C$ is the charge-conjugation matrix and the projector $\mathcal{P}_+$ is applied to isolate the desired positive-parity nucleon states. 
The so-called non-relativistic projector $\Gamma_\text{NR} = (1+\gamma_4)/2$ improves the signal~\cite{Yoon:2016dij}; though more intentional interpolator choices to improve large-momentum signal have been recently developed~\cite{Zhang:2025hyo,Reitinger:2026hta}.
To improve the overlap with boosted nucleon states, we employ Gaussian momentum smearing~\cite{Bali:2016lva} with momentum-smearing parameter $ak=5.6$, which introduces a momentum-dependent phase into the spatial smearing kernel to bias the quark fields toward the target momentum $P_z$, enhancing the signal quality of the corresponding correlation functions up to $P_z = 10 \times 2\pi/L\approx 4.4$ GeV.
Using these momentum-smeared interpolating operators, we construct the corresponding three-point (3pt) correlation functions by inserting a nonlocal gluon operator between the nucleon source and sink. 
These correlators provide access to the gluon structure of the boosted nucleon and are defined as
\begin{multline} \label{eq:3pt-correlator}
    C_{\rm N}^{\rm 3pt} (z, P_z; t_{\rm sep}, t)
    = \\
    \int d^3 y \; e^{-i \mathbf{y}\cdot \mathbf{\hat{z}}P_z} \langle 0 | \chi(\vec{y}; t_{\rm sep}) | O_g(z, t) | \bar{\chi}(\vec{0}; 0) | 0 \rangle,
\end{multline}
where $t_{\rm sep}$ is the source--sink separation and $t$ is the operator insertion time. 
In practice, the gluonic insertion is evaluated through contractions of the nucleon two-point functions with the corresponding gluon-loop contributions. 
The unpolarized gluon PDF operator, first introduced in Ref.~\cite{Balitsky:2019krf}, is defined as 
\begin{multline} \label{eq:O0IB}
    O_g(z,t)
    = 
    F^{ti}(z) W(z,0) F^t_i(0) \\ - F^{ij}(z)W(z,0)F_{ij}(0),
\end{multline}
where $F^{\mu \nu}$ is the gluon field-strength tensor and $W(z,0)$ is a Wilson line connecting the field-strength insertions along the spatial $\hat{z}$-direction. 
The operator indices $(i,j)$ are summed over the relevant spatial directions.
To reduce ultraviolet fluctuations in the gluonic matrix elements, we apply Wilson flow~\cite{Luscher:2010iy} to the gauge links before evaluating the gluon-loop contributions. 
The flowed gauge links $V_{\mu}(x,\tau)$ evolve from the original links $U_\mu(x)$ according to the differential equation
\begin{align}
    & \dot{V}_{\mu}(x,\tau)
    = - g_0^2 \left[ \partial_{x,\mu} S_W(V(\tau)) \right] V_{\mu}(x,\tau), \nonumber
    \\[5pt]
    &V_{\mu}(x,0)=U_{\mu}(x),
\end{align}
where $\tau$ denotes the Wilson flow time, $S_W$ is the Wilson gauge action, and $\partial_{x,\mu}$ is the gauge-covariant derivative with respect to the link $V_\mu(x,\tau)$. 
The flow smooths the gauge fields by suppressing short-distance fluctuations, thereby reducing ultraviolet noise in the gluon-loop contributions. 
We integrate the flow equation numerically using 100 steps at flow time\sout{s} $\tau=3a^2$, denoted as WILSON3. 

\subsection{Analysis Methods}
\label{sec:method}
We extract ground-state matrix elements from the nucleon two-point correlator $C_{\rm 2pt}(P_z;t)$ and the disconnected three-point correlator $C_{\rm 3pt}(z,P_z;t_{\rm sep},t_{\rm ins})$, where $t_{\rm sep}$ is the source--sink separation and $t_{\rm ins}$ the insertion time of the gluon operator. Their spectral decompositions are 
\begin{align} 
  C_{\rm 2pt}(P_z;t) &
  =
  \sum_n |A_n|^2 e^{-E_n t}, \label{eq:c2}\\
  C_{\rm 3pt}(z,P_z;t_{\rm sep},t_{\rm ins}) &= \sum_{n',n} A_{n'} A_n^*
  \langle n'(P_z)|O_g(z)|n(P_z)\rangle \non\\
  &\quad\times e^{-E_{n'} (t_{\rm sep}-t_{\rm ins})}\, e^{-E_n t_{\rm ins}},\label{eq:c3}
\end{align}
with $A_n$ the overlap of the nucleon interpolator with state $n$ of energy $E_n(P_z)$.
The bare ground-state matrix element $M(z,P_z)\equiv\langle 0(P_z)|O_g(z)|0(P_z)\rangle$ represents the target of the correlator anaylsis.
Traditionally, MSULat has extracted these matrix elements by fitting model parameters defined by the two-state truncation of the spectral decompositions in Eqs.~\ref{eq:c2} and~\ref{eq:c3}.
In this study, we use the Lanczos method for the first time and compare with determination of these matrix elements from a two-state fits to validate the newly applied methodology.

\begin{figure*}[tbp]
\centering
\includegraphics[width=0.48\textwidth]{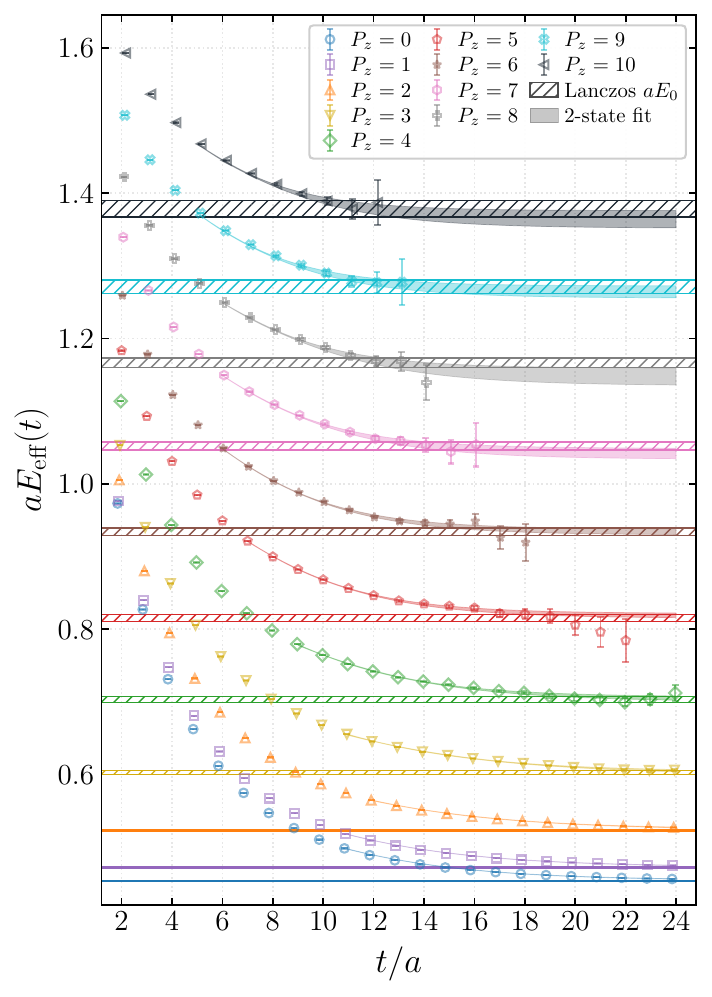}\hfill
\includegraphics[width=0.48\textwidth]{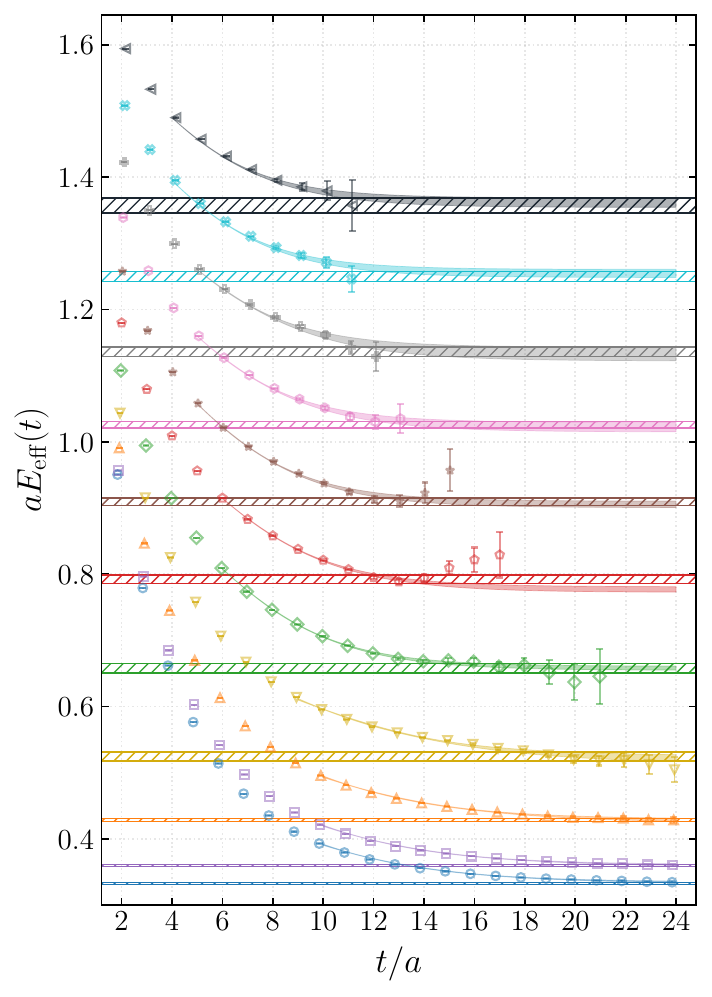}
\caption{
Effective mass $aE_{\rm eff}(t)$ of the nucleon with strange (left) and light (right) valence quarks for boost momenta $aP_z=n_{pz}\,2\pi/L$, $n_{pz}\in[0,10]$. 
Open markers are the jackknife effective-mass data. 
For each momentum, data are shown up to the first time slice whose uncertainty exceeds $0.042$ for clarity.
Hatched horizontal bands show the Lanczos ground-state energy $aE_0$, and shaded bands of the same color show the effective mass reconstructed from the two-state fit on its BAIC-selected window, drawn from $t_{\min}$ onward. 
All bands show $1\sigma$ uncertainties.
}
\label{fig:meff}
\end{figure*}

\paragraph{Two-State Fits.}
As an independent cross-check of the Lanczos spectrum, we also fit $C_{\rm 2pt}(P_z;t)$ directly to the two-state truncation of Eq.~(\ref{eq:c2}) (shaded bands in Fig.~\ref{fig:meff} and open symbols in Fig.~\ref{fig:disp}),
\begin{equation} \label{eq:c2twostate}
  C_{\rm 2pt}^{\rm fit}(P_z;t)
  =
  A_0^2 e^{-E_0 t}+A_1^2 e^{-E_1 t},
\end{equation}
where $E_0$ and $E_1 = E_0 + \Delta E$ are the ground- and excited-state energies, respectively, while $A_0$ and $A_1$ denote the corresponding amplitudes.
The four fit parameters $(A_0,E_0,A_1,\Delta E)$ are treated as free parameters, with the energy gap constrained to $\Delta E\equiv E_1 - E_0 > 0$.
For each momentum $P_z$, we consider candidate fit windows $W=[t_{\min},t_{\max}]$ consisting of contiguous time slices within $[1,t_{\rm snr}]$, with at least eight points. 
Here, $t_{\rm snr}$ denotes the last time slice in the contiguous interval starting at $t=1$ for which the two-point correlator satisfies $|\overline{C}_{\rm 2pt}(P_z;t)|/\sigma_{C_{\rm 2pt}}(P_z;t)\ge 2$. 
Thus, $N_{\max} = t_{\rm snr}$ is the total number of time slices satisfying this signal-to-noise criterion in the initial interval.
Each window is fit by an uncorrelated $\chi^2(W)$ built from the diagonal jackknife uncertainties of $C_{\rm 2pt}(P_z;t)$. 
Fits with $\chi^2(W)/{\rm dof}>3$, or with $E_0$ or $\Delta E$ pinned at the lower bound, are discarded.
The surviving windows are scored with the Bayesian Akaike information criterion (BAIC)~\cite{Neil:2022joj},
\begin{align} 
  {\rm BAIC}(W) &
  =
  \chi^2(W)+2k+2\big[N_{\max}-N_{\rm pts}(W)\big], 
  \label{eq:baic}\\
  w_W &
  =
  \frac{\exp\!\big[-\tfrac{1}{2}{\rm BAIC}(W)\big]}{\sum_{W'}\exp\!\big[-\tfrac{1}{2}{\rm BAIC}(W')\big]},
\end{align}
where $2[N_{\max}-N_{\rm pts}(W)]$ penalizes each time slice excluded from $W$.
We select the highest-BAIC-weight window for which a leave-one-out jackknife refit satisfies the stability criteria: 
\begin{equation}
    \begin{aligned}
    \frac{\sigma(E_0)}{|E_0|} &\le 0.02, &
    \frac{\sigma(A_0)}{|A_0|} &\le 0.10, &
    \frac{\sigma(\Delta E)}{|\Delta E|} &\le 1.
    \end{aligned}
\end{equation}
%
We then quote the jackknife mean and uncertainty of the fit parameters obtained from the selected window.

\paragraph{Lanczos.}
We apply Lanczos as described in Refs.~\cite{Wagman:2024rid,Hackett:2024xnx,Hackett:2024nbe} to the two- and three-point correlators. 
Regarding the two-point correlator, we start the Krylov space at $t_0=2a$, discarding the contact-dominated time slices $t/a\in\{0,1\}$, giving a maximal Krylov dimension of $m_{\max}=(n_t - t_0/a) / 2 = 17$. 
Spurious Ritz values are removed using the ZCW overlap test of Ref.~\cite{Hackett:2024nbe}. 
The threshold $\varepsilon_{\rm ZCW}$ is defined as the minimum ZCW overlap at the last iteration for which at least $90\%$ of bootstrap samples have entirely real spectra, divided by $F_{\rm ZCW} = 10$.
Surviving states are then ordered by energy. 
For the three-point function, we truncate the filtered Ritz basis to its two lowest states and extract $M(z,P_z)$ by projecting $C_{\rm 3pt}$ onto the ground state within this two-state subspace. 
Requiring both the source--operator and operator--sink separations to be at least $2a$ restricts the total source--sink separation to $t_{\rm sep}/a\in[4,34]$.

All Lanczos estimates are computed on correlated nested bootstrap resamplings of the two- and three-point functions with $N_{\rm b}^{\rm inner}=N_{\rm b}^{\rm outer}=200$ samples. 
Central values are nested medians and uncertainties are central 68\% bootstrap confidence intervals~\cite{Wagman:2024rid,Hackett:2024xnx}. 
The ground-state energies are quoted over the final $N_m=3$ iterations, while the matrix elements are quoted at $m=17$.


\subsection{Two-point results}\label{sec:2pt}
We first check the consistency of the Lanczos and two-state fit extractions of ground state energy from the 2pt correlators by inspecting the effective mass 
\begin{equation}\label{eq:eff_mass}
    aE_{\rm eff}(t)
    =
    \ln \left[\frac{C_{\rm 2pt}(t)}{C_{\rm 2pt}(t+1)}\right] 
    \xrightarrow[]{t\rightarrow\infty} 
    aE(P_z),
\end{equation}
where $E(P_z)$ is the ground state energy of the nucleon with momentum $P_z$.
Fig.~\ref{fig:meff} shows the effective masses for the nucleon constructed with both light (right) and strange (left) valence quarks at lattice momenta corresponding to $a P_z = 2\pi n_{pz} /L$ for integer $n_{pz} \in [0,10]$. 
At early times, $aE_{\rm eff}(t)$ decreases as excited-state contributions decay, eventually approaching a plateau consistent with the Lanczos ground-state energy at each momentum.

As $P_z$ increases and the signal-to-noise ratio decays earlier in Euclidean time, the BAIC-selected two-state fit window shifts toward smaller times: from $[11,35]$ to $[5,15]$ for the strange nucleon and from $[10,35]$ to $[4,14]$ for the light nucleon as $P_z$ increases from $0$ to $10$.
In contrast, Lanczos uses the full correlator from $t_0=2a$ at every momentum, with no fit range to choose.
The two determinations of $aE_0(P_z)$ have comparable statistical uncertainties and agree within $1.5\sigma$ at all eleven momenta for the strange nucleon. 
For the light nucleon, they agree within $0.6\sigma$ at every momentum, except $P_z=5$, where the Lanczos value $aE_0=0.793(6)$ lies $2.1\sigma$ above the two-state fit value $0.777(4)$.
At zero momentum, the Lanczos first-excited-state gaps are consistent with the two-state-fit results: $a\Delta E = 0.225(13)$ versus $a\Delta E = 0.247(4)$ for the strange nucleon, and $0.258(18)$ versus $0.276(14)$ for the light nucleon.

\begin{figure*}[tbp]
\centering
\includegraphics[width=0.48\textwidth]{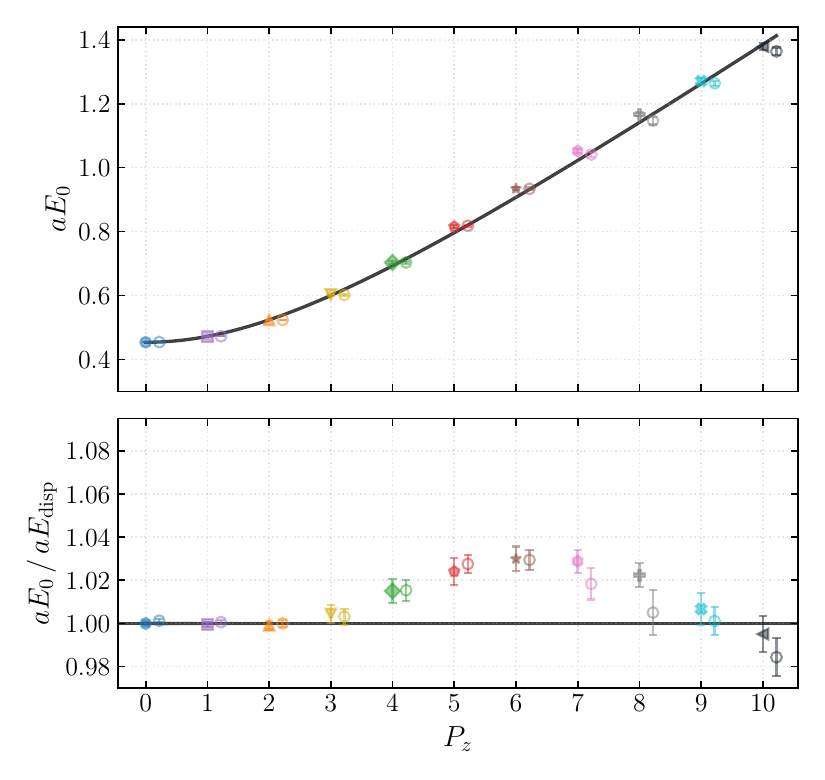}\hfill
\includegraphics[width=0.48\textwidth]{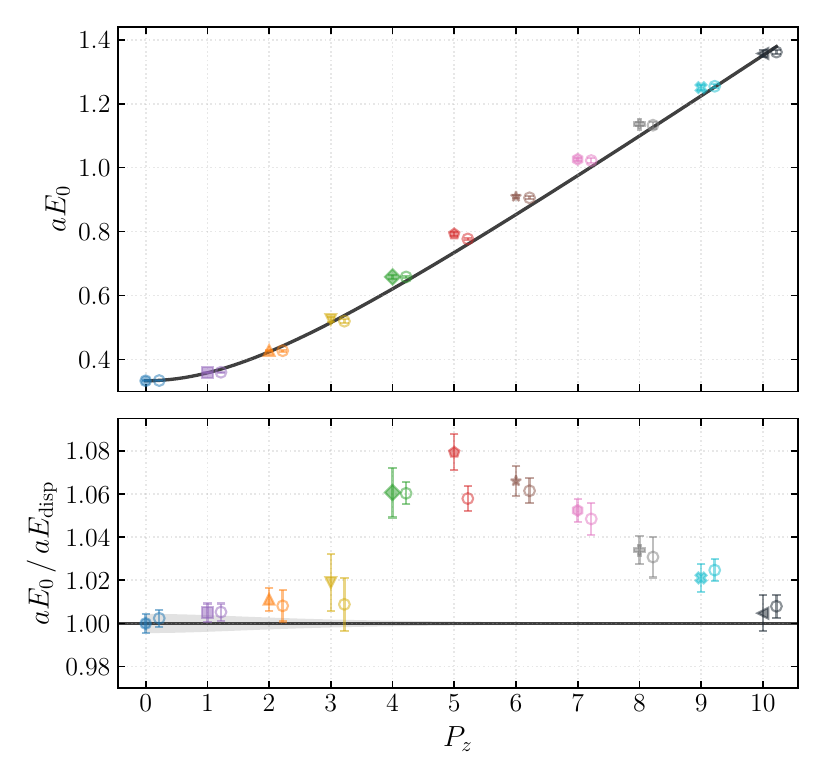}
\caption{
Ground-state energy $aE_0(P_z)$ versus boost momentum $a P_z=n_z\,2\pi/L$, $n_z\in[0,10]$, for the nucleon with strange (left) and light (right) valence quarks. 
At each momentum, filled markers show the Lanczos result and open markers, offset to the right for clarity, show the two-state fit. 
Colors and marker follow Fig.~\ref{fig:meff}. 
The solid curve is the continuum dispersion relation $aE_{\rm disp}(P_z)=\sqrt{(aM)^2+(aP_z)^2}$ evaluated at the Lanczos rest mass $aM=aE_0(0)$. 
The lower panels show the ratio $aE_0(P_z)/aE_{\rm disp}(P_z)$ for both extractions, with the gray band indicating the uncertainty in $aE_{\rm disp}$ propagated from $aM$}
\label{fig:disp}
\end{figure*}

Figure~\ref{fig:disp} compares the ground-state energies from both extractions against the continuum dispersion curve $E_{\rm disp}(P_z) = \sqrt{(M)^2+( P_z)^2}$ for both the strange (left) and light (right) nucleons, evaluated using Lanczos rest masses $aM = 0.45349(49)$ and $0.3331(15)$ respectively.
For the strange nucleon, both extractions follow the continuum curve over the full range of momenta: the ratio $E_0(P_z)/E_{\rm disp}(P_z)$ remains within $3\%$ of unity for both Lanczos and the two-state fits.
The light nucleon deviates further from the continuum curve. 
Both extractions lie $6$--$8\%$ above it at $P_z=4$--$6$, with the largest ratios $1.079(8)$ for Lanczos at $P_z=5$ and $1.061(6)$ for the two-state fit at $P_z=6$, before returning within $1\%$ of unity at $P_z=10$.
This larger deviation, which is common to both extraction methods, may reflect lattice artifacts such as discretization or finite volume effects. 
We do not attempt to disentangle these in the present work. 
In either case, the deviation from the continuum dispersion relation remains at the level of a few percent over the full range of momenta, indicating that such effects are small at boost.

\subsection{Three-Point Results}
\label{sec:3pt}
\begin{figure*}[tbp]
\centering
\includegraphics[width=0.48\textwidth]{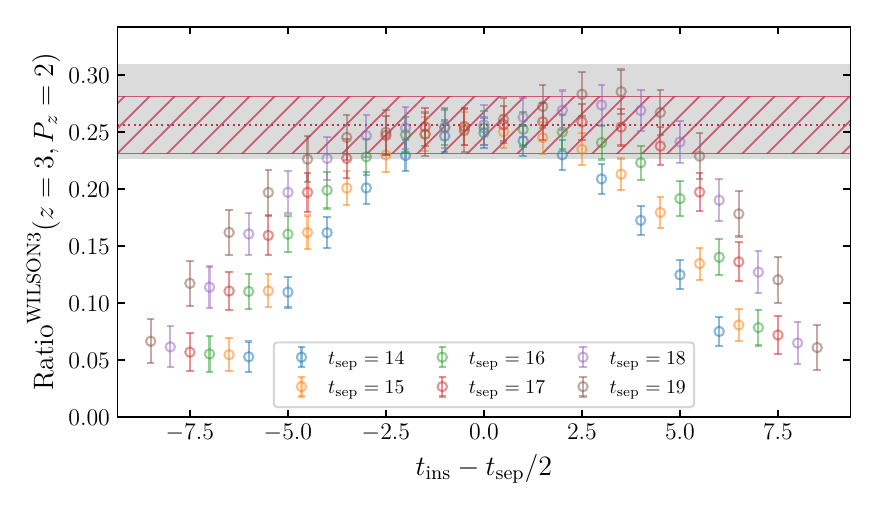}
\includegraphics[width=0.48\textwidth]{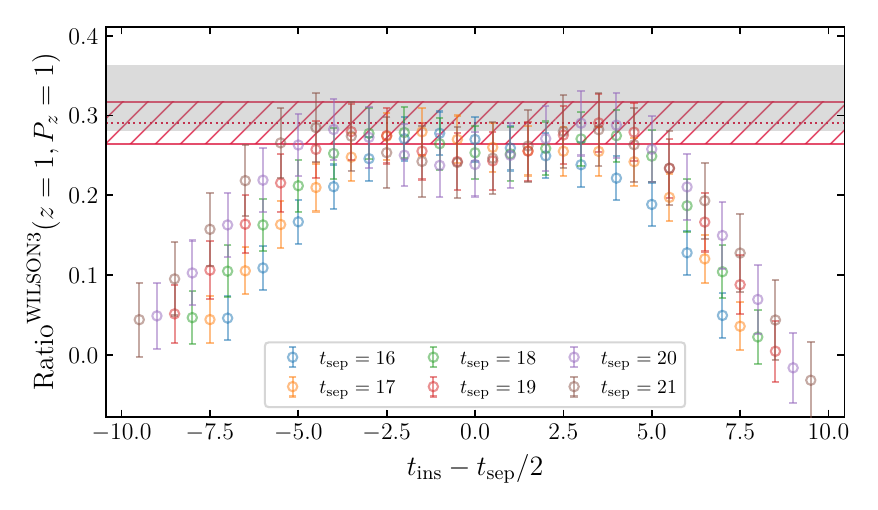}
\includegraphics[width=0.48\textwidth]{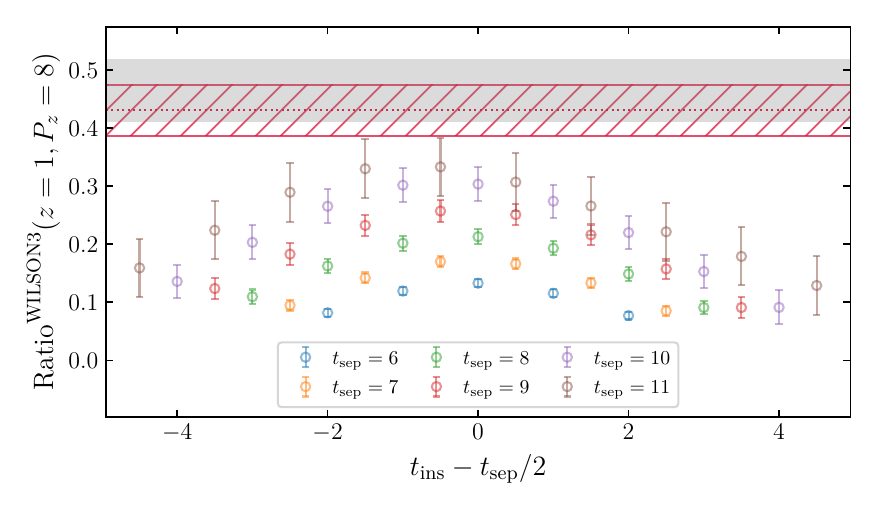}
\includegraphics[width=0.48\textwidth]{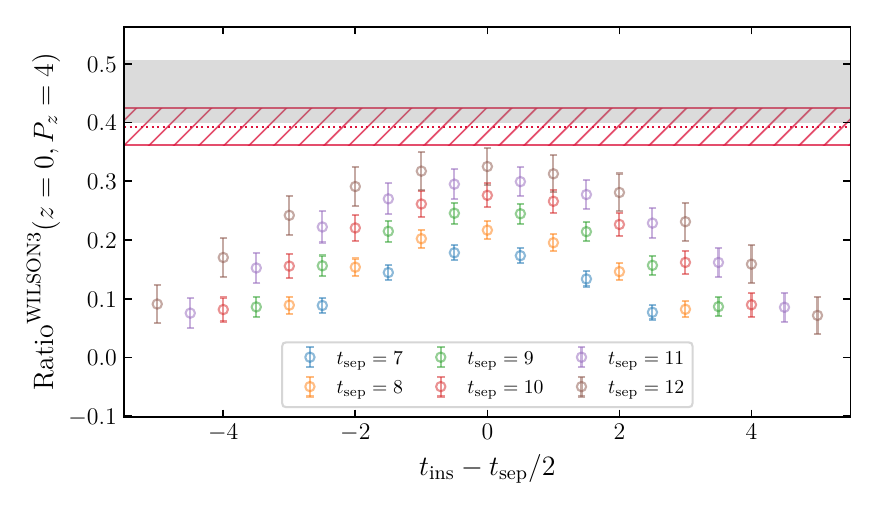}
\caption{
Ratio $R(z,P_z;t_{\rm sep},t_{\rm ins})$ of the three- and two-point correlators, Eq.~\eqref{eq:2pt-3pt-correlator-ratio}, versus $(t_{\rm ins}-t_{\rm sep}/2)/a$ for the strange (left) and light (right) nucleons. 
The panels show $(P_z,z)=(2,3)$ (top left), $(8,1)$ (bottom left), $(1,1)$ (top right), and $(4,0)$ (bottom right), with $P_z$ in units of $2\pi /L$ and $z$ in units of $a$. 
Open circles are the jackknife estimates of the ratio at six selected source--sink separations $t_{\rm sep}/a$ (see legends), shown for$1\le t_{\rm ins}/a\le t_{\rm sep}/a-1$. 
The hatched crimson band with dotted central line is the Lanczos ground-state matrix element, and the solid gray band is the model-averaged two-state fit result. 
Both bands show the total uncertainty.
}
\label{fig:ratio}
\end{figure*}

Next, we can compare the two-state and Lanczos extraction of matrix elements from the three-point data.
A common way to visualize the quality of this extraction is to look at the ratio between the three- and two-point data
\begin{multline}\label{eq:2pt-3pt-correlator-ratio}
    R(z,P_z;t_\text{sep}, t_\text{ins}) 
    \\ = \frac{C_{\text{3pt}} (z, P_z; t_\text{sep}, t_\text{ins})}{C_{\text{2pt}} (P_z; t_\text{sep})}
    \\ \approx M(z,P_z) + \mathcal{O}(e^{-\Delta E t_\text{sep}}).
\end{multline}
Figure~\ref{fig:ratio} displays this ratio for representative values of $z$ and $P_z$ for the strange (left column) and light (right column) nucleons.
At low boosts (top row), the ratios exhibit a plateau near $t_{\rm ins}=t_{\rm sep}/2$ for the displayed source--sink separations, and the Lanczos and two-state-fit results agree within $1\sigma$. 
For the strange nucleon at $(P_z,z)=(2,3)$, we obtain $M=0.256(25)$ and $0.268(42)$, respectively, while for the light nucleon at $(P_z,z)=(1,1)$, we obtain $M=0.291(26)$ and $0.322(41)$.
At higher boost (bottom row), the signal-to-noise ratio decays much earlier, restricting the usable separation times to $t_{\rm sep}/a\lesssim 12$. 
Over this range, the ratio continues rising with $t_{\rm sep}$, with even the largest displayed separation lies $20$--$30\%$ below the extracted matrix element, so both methods must extrapolate to the ground state.
The two extractions again agree within $1\sigma$, with $M=0.430(44)$ and $0.465(54)$ for the strange nucleon at $(8,1)$ and $0.393(32)$ and $0.453(54)$ for the light nucleon at $(4,0)$.

\section{Results and Discussions}
\label{sec:results}
\begin{figure*}[tbph]
    \centering
    \includegraphics[width=0.75\textwidth]{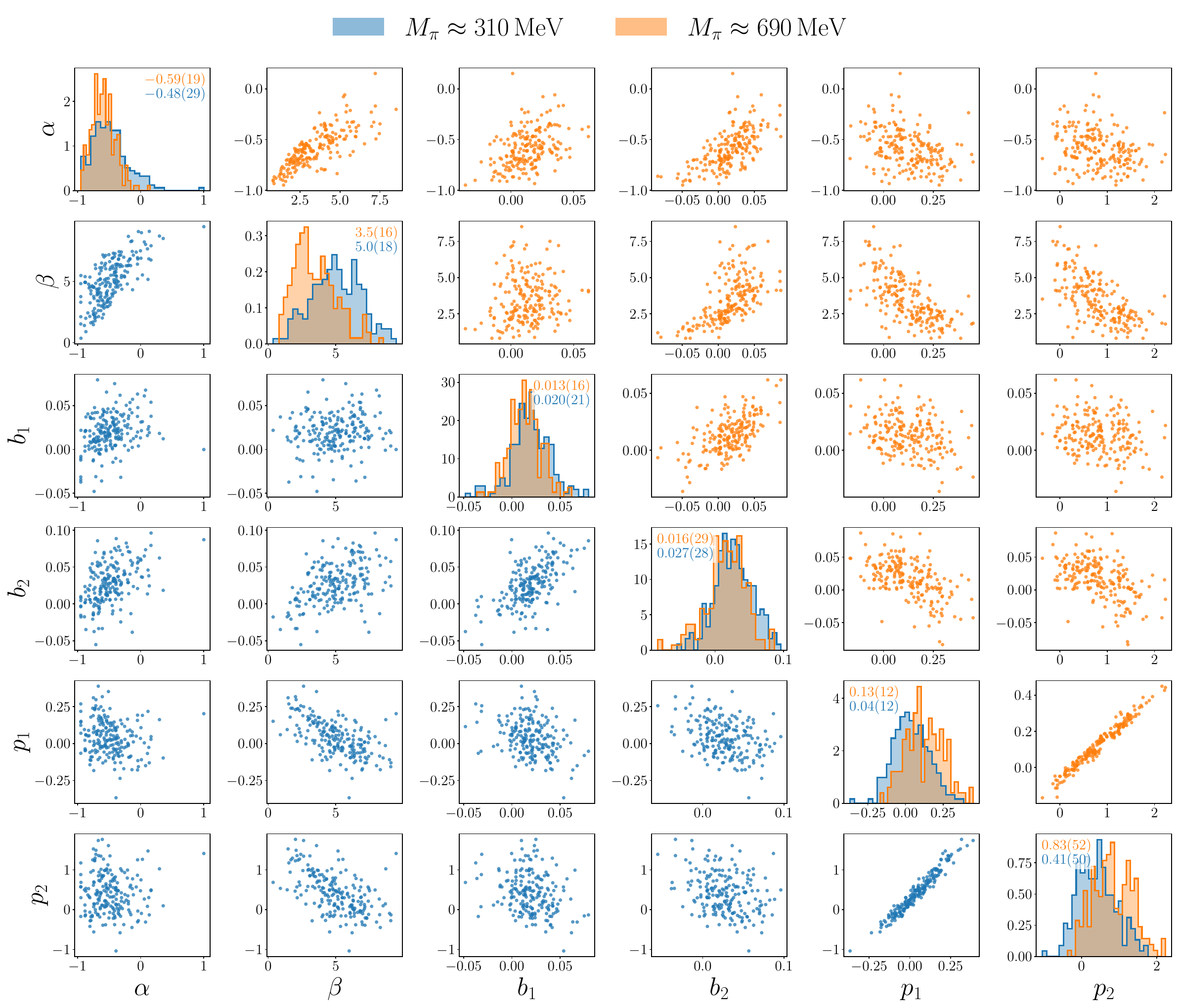}
    \caption{
    Bootstrap distributions (diagonal panels) and pairwise correlations (off-diagonal panels) of the fitted parameters ${\alpha,\beta,b_1,b_2,p_1,p_2}$ for the $M_\pi \approx 310$ MeV (lower triangle) and $M_\pi \approx 690$ MeV (upper triangle) ensembles. 
    The diagonal panels show the bootstrap distributions of the individual fit parameters, with their means and statistical uncertainties annotated in orange and blue text for the respective pion masses, while the off-diagonal panels show the joint bootstrap distributions of parameter pairs, revealing correlations and anticorrelations arising from the fit. 
    Comparing the lower and upper triangles highlights differences in the parameter uncertainties and correlation structure between the two pion-mass ensembles.
    }
    \label{fig:param-correlations}
\end{figure*}

\begin{figure*}[th]
    \centering
    \includegraphics[width=0.45\textwidth]{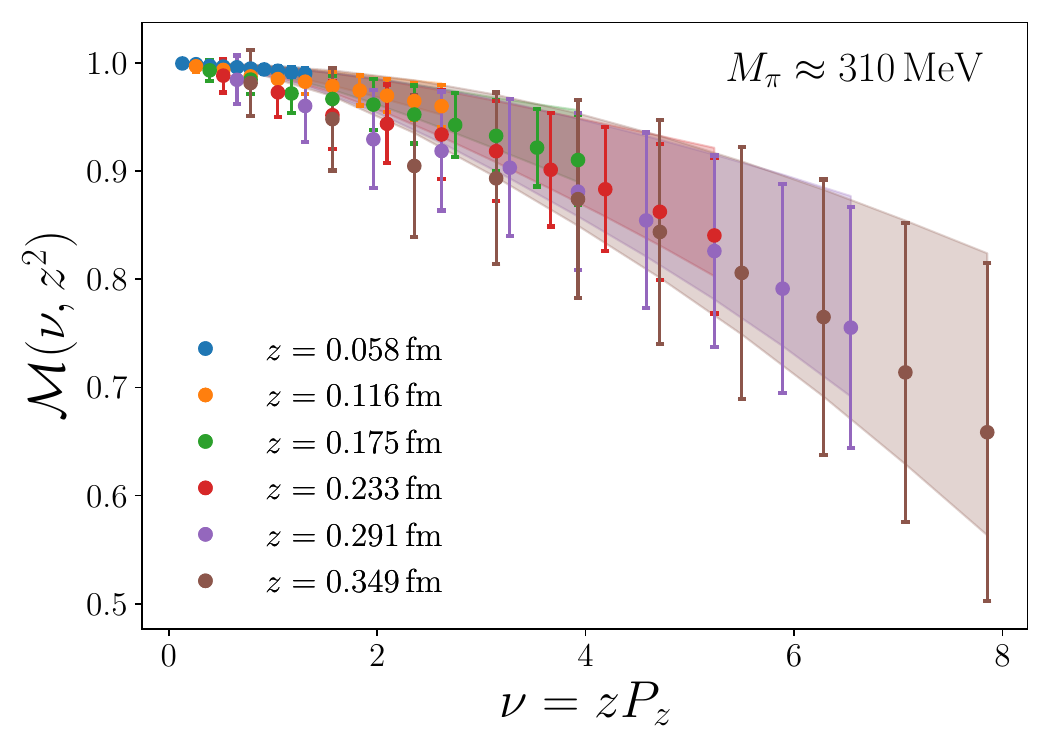}
    \includegraphics[width=0.45\textwidth]{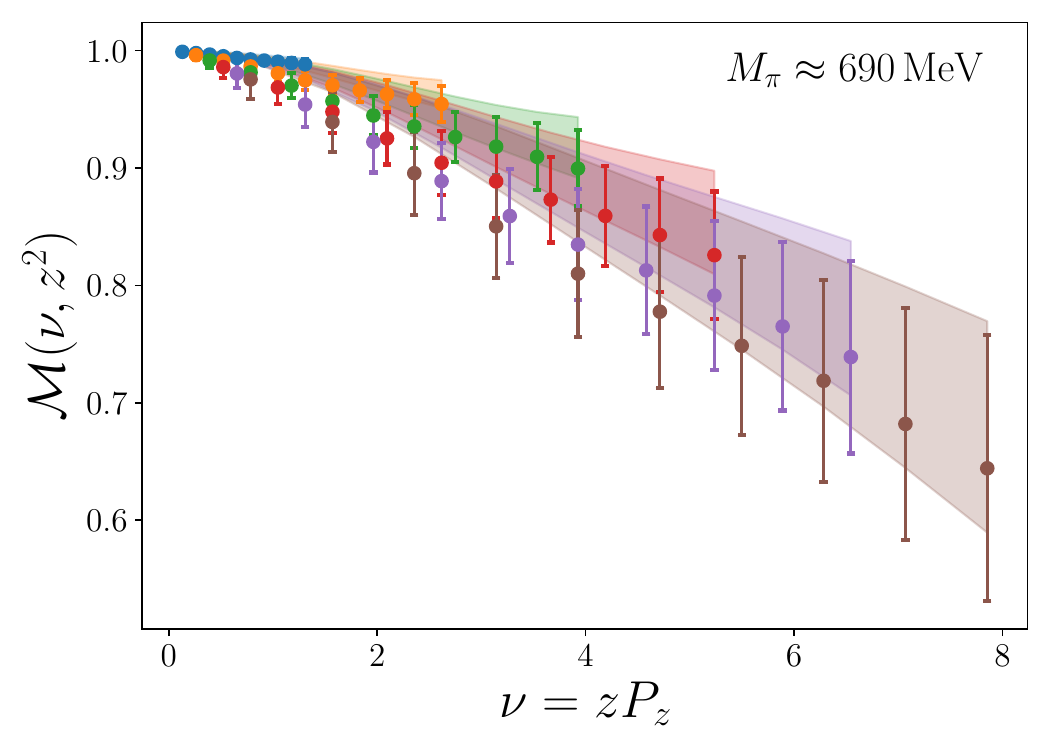}
    \caption{
    Lattice RpITD data and reconstructed model predictions as functions of the Ioffe time $\nu=zP_z$ for different spatial separations $z$, shown for the light (left) and strange (right) nucleons. 
    The lattice data are obtained at discrete values of momentum $P_z$ and spatial separation $z$, while the reconstructed RpITDs are evaluated using the fitted model parameters to describe the observed dependence on $\nu$ and $z$. 
    Error bars denote the uncertainties in the lattice data, while shaded bands indicate the bootstrap uncertainties associated with the reconstructed RpITDs. 
    The comparison between the lattice data and reconstructed data illustrates the quality of the fit across the available kinematic range and allows the consistency of the reconstruction to be assessed for both pion masses and different spatial separations.
    }
    \label{fig:reconstructed-RpITDs}
\end{figure*}

The bare matrix elements extracted in the previous section are multiplicatively renormalizable as~\cite{Zhang:2018diq}
\begin{equation}
    M^B(z,P_z) 
    =
    Z_L(1/a) e^{-|z|\delta m}M^R(z,P_z),
\end{equation}
where $Z_L(1/a)$ renormalizes the logarithmic UV divergences the lattice cutoff $1/a$, and the exponential term renormalizes the linear divergence associated with the Wilson line self-energy, $\delta m$.
In the pseudo-PDF method, the most convenient renormalization scheme is to construct the reduced pseudo-Ioffe time distribution (RpITD):
\begin{equation}\label{eq:RpITD}
    \mathcal{M}(\nu=zP_z,z^2) = \frac{M^B(0,0)}{M^B(0,P_z)}\frac{M^B(z,P_z)}{M^B(z,0)}.
\end{equation}
The term $M^B(0,0)/M^B(0,P_z)$ removes kinematic factors coming from state normalization and the gluon operator definition, while division by $M^B(z,0)$ cancels the $z$-dependent renormalization factors.
The RpITD is then a function of the Lorentz invariant Ioffe time $\nu = zP_z$ and separation distance squared, $z^2$.
The pseudo-PDF methodology is based on matching between the pseudo-ITD and the light-cone ITD, which is related to the gluon PDF just through a Fourier transform.
In practice, the two steps of this matching can be written as a single relation~\cite{Balitsky:2019krf, Balitsky:2021qsr}:
\begin{eqnarray} \label{eq:matching}
     && \hspace*{-0.5cm}
\mathcal{M}(\nu, z^2)  = 
     \notag\\ 
     && \hspace*{-0.5cm}
 \int_0^1 \dd x \frac{xg(x)}{\langle x_g \rangle}
  \left[R_{gg}(x\nu, z^2) 
      + R_r(x\nu, z^2) \frac{\langle x_\Sigma \rangle}{\langle x_g \rangle} \right]  
      \notag\\
     && \hspace*{-0.5cm} + \int_0^1 \dd x \frac{x\Sigma(x)}{\langle x_g \rangle} R_{gq}(x\nu, z^2) + \mathcal{O}(m^2z^2, z^2 \Lambda_\text{QCD}^2),
\end{eqnarray}
where $g(x)$ and $\Sigma(x)$ are the $\overline{\text{MS}}$ scheme gluon and quark singlet PDFs respectively.
The first moments of these two distributions are represented by $\langle x_g\rangle$ and $\langle x_\Sigma\rangle$.
Higher twist effects come into play in this matching at $\mathcal{O}(m^2z^2, z^2 \Lambda_\text{QCD}^2)$, and are suppressed here by ensuring that we use separation distances within the perturbative range $z \lesssim 0.3-0.4$ fm.
The terms $R_{gg}$, $R_{r}$, and $R_{gq}$ are perturbatively calculable matching kernels.
The first kernel $R_{gg}$ is a Fourier transform at leading order and contains next-to-leading order (NLO) contributions from gluon diagrams.
The $R_r$ and $R_{gq}$ kernels contribute at NLO from quark mixing terms, and both are needed to maintain the exact normalization of $\mathcal{M}(\nu=0, z^2)  = 1$.
The quark mixing contribution has been shown to be negligible compared to current statistical and systematic uncertainties in the lattice data~\cite{Fan:2021bcr, Fan:2022kcb, Delmar:2023agv}, as the mixing is suppressed at NLO by the ratio of color factors $C_F/C_A$.
Hence, we will ignore this term for this study.
Instead, we focus on reducing higher-twist terms and discretization effects by introducing the matching relation as 
\begin{eqnarray} \label{eq:matching2}
      \hspace*{-0.5cm}
      \mathcal{M}(\nu, z^2)  
      =
      \int_0^1 \dd x R_{gg}(x\nu, z^2) \frac{xg(x)}{\langle x_g \rangle} &
      \notag\\ 
      \hspace{.5cm} + z^2 B_1(\nu) + \frac{1}{z}P_1(\nu).&
\end{eqnarray}
The terms $z^2 B_1(\nu)$ and $(1/z)P_1(\nu)$ are designed to control systematic higher-twist and discretization effects~\cite{Karpie:2021pap, JeffersonLabAngularMomentumJAM:2022aix}, respectively, and are parametrized by
\begin{align} \label{eq:sys_param}
    B_1(\nu) &= \sum_{n=1}^N b_n\, \sigma_{0,n}(\nu), \quad
    P_1(\nu)  = \sum_{n=1}^N p_n\, \sigma_{0,n}(\nu), 
\end{align}
where $\sigma_{0,n}$ are weighted cosine transforms of Jacobi polynomials, shifted to the domain $0\leq x \leq 1$,
\begin{equation} \label{eq:sigma_def}
    \sigma_{0,n}(\nu) = \int_0^1 \dd{x}\, \cos(\nu x)\, x^a (1-x)^b\, J_n^{(a,b)}(x).
\end{equation}
In this work, we retain only the first two basis functions, $N=2$, and impose $\sigma_{0,n}(0)=0$ to ensure that the systematic corrections preserve the normalization of the reduced pseudo-Ioffe-time distribution (RpITD). 

We construct the gluon PDF using the parameterization
\begin{equation}
    \frac{xg(x)}{\langle x \rangle_g}
    =
    \frac{x^\alpha (1-x)^\beta}{B(\alpha + 1, \beta + 1)}
\end{equation}
where $\langle x\rangle_g$ is the gluon momentum fraction, $B(\alpha+1,\beta+1)$ is the Euler beta function, and the parameters $\alpha$ and $\beta$ control the small- and large-$x$ behavior of the distribution, respectively. 
This parameterization determines the leading-twist contribution RpITD, which is fitted simultaneously with the higher-twist and discretization corrections defined in Eq.~\ref{eq:sys_param}. 
The resulting model depends on the six fit parameters $\boldsymbol{\theta} = \{\alpha, \beta, b_1, b_1, p_1, p_2\}$, which are determined through a correlated fit to the RpITD data. 
Specifically, we minimize the correlated chi-squared,
\begin{multline} \label{eq:correlated-chi2}
    \chi^2(\boldsymbol{\theta})
    =
    \left[\mathcal{M}(\nu, z^2) - \mathcal{M}(\nu, z^2; \boldsymbol{\theta})\right]^T \mathbf{C}^{-1} \\ \times \left[\mathcal{M}(\nu, z^2) -\mathcal{M}(\nu, z^2; \boldsymbol{\theta})\right],
\end{multline}
where $\mathcal{M}(\nu, z^2)$ and $\mathcal{M}(\nu, z^2; \boldsymbol{\theta})$ denote the vectors of measured and modeled RpITD values, respectively. 
The covariance matrix $\mathbf{C}$ encodes correlations among the data points and is used to account for these correlations in determining the best-fit parameters and their uncertainties. 
We repeat the fitting procedure for the bootstrap samples to estimate the uncertainties and correlations of the extracted parameters.

To investigate the constraints imposed by the correlated fits, we examine the joint distributions of the fitted parameters across the bootstrap samples. 
Figure~\ref{fig:param-correlations} presents the resulting parameter distributions and pairwise correlations for the light and strange nucleons. 
The diagonal panels show the marginal bootstrap distributions of the six fit parameters, while the off-diagonal panels display their pairwise relationships. 
For clarity, the lower triangle shows correlations for the light nucleon (blue points), and the upper triangle shows those for the strange nucleon (orange points). 
These distributions provide insight into the uncertainties of the extracted parameters and the correlations between the PDF shape and the systematic corrections.
The marginal distributions exhibit broadly similar behavior across the two pion masses, suggesting that the fit parameters are constrained in a qualitatively consistent manner for both ensembles.
The pairwise distributions reveal a pronounced correlation between $\alpha$ and $\beta$ which reflects the ill-conditioned nature of the inverse problem, whereby different combinations of endpoint exponents can produce similar predictions for the lattice data, leaving the individual parameters less constrained.
An even stronger correlation is observed between $p_1$ and $p_2$, which govern the discretization correction. 
Since both coefficients enter through a shared $\nu$-dependent contribution proportional to $a/|z|$, their effects can partially compensate for one another over the fitted kinematic range. 
These correlations illustrate how the lattice data constrain combinations of fit parameters rather than necessarily determining each parameter independently. 
Accounting for these dependencies is therefore essential when propagating the bootstrap uncertainties to the reconstructed RpITDs and gluon PDF.
Overall, we expect better control of the PDF parameters to come as the precision fo large-$\nu$ data is improved.
The $p_i$ parameters could be better constrained by the inclusion of multiple lattice spacings  to uniquely separate $a/z$ discretization effects from the $\nu$-dependence of $P_1(\nu)$.

\begin{figure*}[th]
    \centering
    \includegraphics[width=0.45\textwidth]{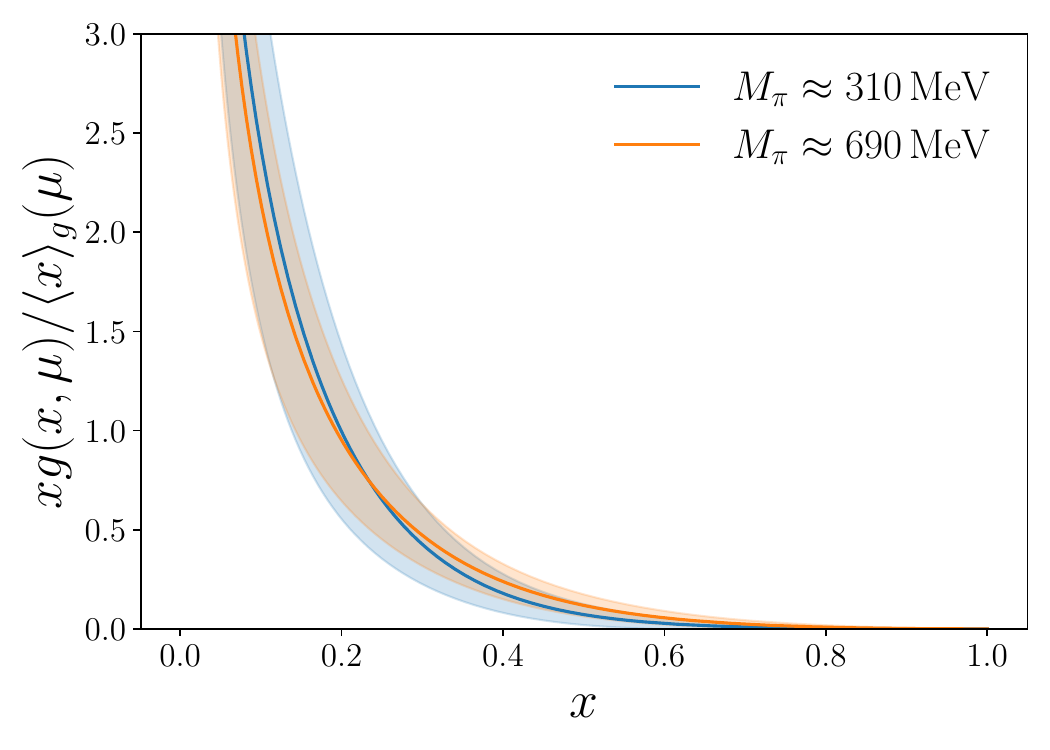}
    \includegraphics[width=0.45\textwidth]{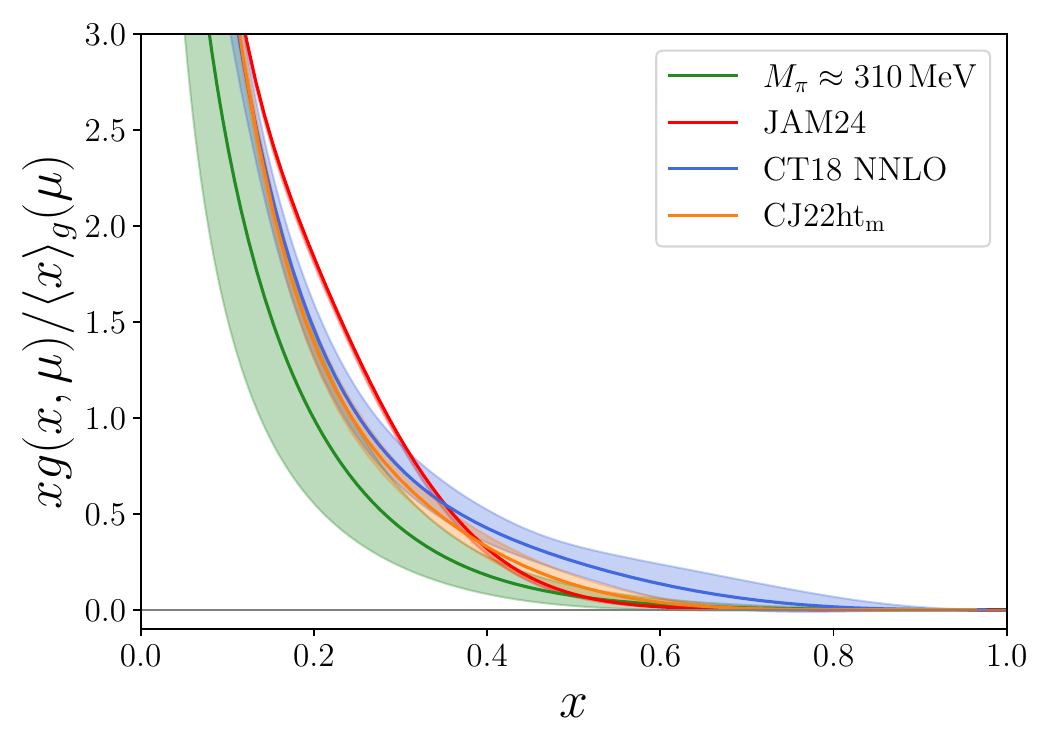}
    \caption{
    Unpolarized nucleon gluon PDFs extracted from the lattice RpITD data. 
    (Left) Comparison of the extracted gluon PDFs for the two pion-mass ensembles, $M_\pi\approx310$ and $690$ MeV, illustrating the pion-mass dependence of the reconstructed distributions and their associated uncertainties. 
    (Right) The gluon PDF extracted from the $M_\pi\approx310$ MeV ensemble compared with the phenomenological global-fit results CJ22~\cite{Cerutti:2025yji}, JAM24~\cite{Anderson:2024evk}, and CT18~\cite{Hou:2019efy}. 
    The comparisons assess the consistency between the lattice extraction and phenomenological determinations across the momentum-fraction range $x$. 
    Differences among the distributions highlight the remaining uncertainties in the gluon PDF, particularly in regions where the RpITD data provide limited sensitivity.
    }
    \label{fig:final-pdfs}
\end{figure*}

Figure~\ref{fig:reconstructed-RpITDs} compares the lattice RpITD data with the reconstructed model predictions as functions of the Ioffe time for the two pion-mass ensembles, with different colors denoting the spatial separations $z$.
The reconstructed predictions incorporate both the leading-twist gluon PDF and the higher-twist and discretization corrections, with the shaded bands representing the uncertainties propagated from the bootstrap distributions of the fitted parameters. 
This comparison allows us to assess how well the fitted model describes the lattice data across the available kinematic range.
For the lighter-pion ensemble ($M_\pi\approx310$ MeV), the reconstructed predictions generally describe the lattice data well within their $1\sigma$ uncertainties, particularly at larger momenta $P_z$. 
However, at smaller momenta, the lattice RpITDs exhibit a more rapid decay than the reconstructed predictions. 
Similar behavior has been reported in previous gluon PDF studies~\cite{NieMiera:2025inn,Good:2024iur,Kriesten:2025gti}, where the lowest-momentum data were found to deviate from the expected matching behavior. 
The overall fit quality remains reasonable, with mean reduced chi-squared values of $\chi^2/\mathrm{d.o.f.}=1.128(228)$ and $1.067(206)$ for the strange and light nucleons, respectively. 
Thus, the current statistical precision is insufficient to determine whether this difference reflects a systematic deviation from the fitted model.
The heavier-pion ensemble ($M_\pi\approx690$ MeV) shows broadly similar agreement, but the reconstructed bands tend to lie above the lattice data at larger separations and smaller momenta. 
Overall, the reconstructed RpITDs provide a comparison between the fitted model and the lattice data, illustrating both the general level of agreement and the variations observed across the available momenta and spatial separations.

Finally, we present the extracted unpolarized nucleon gluon PDFs for the two valence pion-masses in left panel of Fig.~\ref{fig:final-pdfs}. 
Comparing the PDFs at $M_\pi\approx310$ and $690$ MeV, we observe statistically consistent distributions across $x$, as expected from the extracted parameters described earlier.
Minimal pion mass dependence in the PDFs are consistent with observations from earlier psuedo-PDF studies~\cite{Fan:2020cpa,Fan:2021bcr,Fan:2022kcb} and recent LaMET extractions~\cite{Good:2024iur,Good:2025daz,ChenChen:2025amm, NieMiera:2025vcx}.
Nonetheless, lighter pion masses, should be explored to ensure that this trend is consistent closer to the physical pion mass.
In the right panel of Fig.~\ref{fig:final-pdfs}, we show the lightest pion mass result compared to select phenomenological results: CJ22~\cite{Cerutti:2025yji}, JAM24~\cite{Anderson:2024evk}, and CT18~\cite{Hou:2019efy}. 
Our results agree with different global fit behaviors at various ranges of $x$.
At $x\leq 0.2$, the phenomenological results sit right at the $1\sigma$ boundary of the lattice results, but we do not expect the RpITD to have much sensitivity the PDF in this range~\cite{Dutrieux:2025jed}.
In particular, our extraction supports the highly suppressed gluon distribution at $x\geq0.5$ of the JAM24 and CJ22 global fits, while presenting tension of greater than $1\sigma$ with the CT18 extraction in the range $0.3 \lesssim x \lesssim 0.6$.
Appreciating that there is still much flexibility in the correlated values of the $\alpha$ and $\beta$ parameters presented in Fig.~\ref{fig:param-correlations}, it seems likely that inclusion of the RpITD data from this analysis into global fits could produce PDFs with good agreement both the experimental and lattice data, as was seen from the pion PDF study presented in Ref.~\cite{Good:2025nny}.

\section{Conclusion and Outlook}\label{sec:conclusion}
We have presented the most systematically controlled extraction of the nucleon gluon PDF using the pseudo-PDF methodology to date. 
Our analysis emphasizes the use of a finer lattice spacing than in previous studies, improved control of excited-state contamination through the recently developed Lanczos-based matrix element analysis~\cite{Hackett:2024nbe}, and explicit parameterizations of systematic effects in the PDF extraction~\cite{Karpie:2021pap, JeffersonLabAngularMomentumJAM:2022aix}.

We compared a rigorous implementation of our previous two-state fit approach with the Lanczos method, finding reasonable agreement between the two. 
The Lanczos method, however, exhibits substantially reduced sensitivity to analysis choices, particularly the selection of fit windows. 
The fine lattice spacing of $a\approx 0.06$ fm enabled the use of six discrete separation distances within the perturbative region $z\lesssim 0.3$--$0.4$ fm, together with ten nucleon momenta reaching approximately $4.4$ GeV. 
These data provide constraints on the gluon PDF at valence pion masses $M\_\pi\approx 310$ and $690$ MeV.

We explicitly parameterized higher-twist contributions and discretization effects and examined correlations among the resulting fit parameters, finding that the data constrain certain correlated combinations of PDF parameters. We observe minimal dependence on the valence pion mass and identify nonzero discretization effects that can be accounted for within our fitting framework. Further control over these effects could be achieved through simultaneous fits to data from multiple lattice spacings. 
The resulting range of PDF parameterizations describes the reduced RpITD data well and exhibits reasonable agreement with phenomenological extractions over various regions of $x$. In particular, we find a strongly suppressed gluon PDF at large $x$.

The broad range of PDF parameterizations compatible with our lattice data, together with the improved control of systematic effects, suggests that these RpITD results could provide useful constraints in global PDF analyses. 
In particular, a simultaneous fit to the lattice RpITD and experimental data could test their mutual consistency and further constrain the gluon PDF, analogous to the joint analysis performed for pion PDFs in Ref.~\cite{Good:2025nny}.

\section*{Acknowledgments}
We thank the MILC Collaboration for sharing the lattices used to perform this study.
The LQCD calculations were performed using the Chroma software suite~\cite{Edwards:2004sx}.
This research used resources of the National Energy Research Scientific Computing Center, a DOE Office of Science User Facility supported by the Office of Science of the U.S. Department of Energy under Contract No. DE-AC02-05CH11231 through ERCAP;
facilities of the USQCD Collaboration, which are funded by the Office of Science of the U.S. Department of Energy,
and supported in part by Michigan State University through computational resources provided by the Institute for Cyber-Enabled Research (iCER).
The work of AN is partially supported by U.S. Department of Energy, Office of Science, under grant DE-SC0024053 ``High Energy Physics Computing Traineeship for Lattice Gauge Theory''.
KE is partly supported by the Early-Start Fellowship from the College of Natural Science at Michigan State University.
The work of WG and HL is partially supported by the US National Science Foundation under grant PHY~2209424 and 2514533.
%
\bibliography{refs.bib}

\end{document}